\documentclass[journal]{IEEEtran}
\usepackage{algorithm,algorithmic}
\usepackage[subtle]{savetrees}
\usepackage{verbatim}
\usepackage{graphicx}
\usepackage{subcaption}
\usepackage{amsmath,amssymb,amsfonts,bm}
\usepackage{epsf}
\usepackage{tikz}
\usepackage{psfrag}
\usepackage{pstool}
\usepackage{epstopdf}
\usepackage{booktabs}
\usepackage{stfloats}
\usepackage{url}
\usepackage[dvips]{epsfig}
\usepackage{amsthm}
\usepackage{color}
\usepackage[usenames,dvipsnames]{pstricks}
\usepackage{lettrine}
\usepackage[dvips]{epsfig}
\usepackage{pst-grad} 
\usepackage{pst-plot} 
\usepackage{epsfig}
\usepackage{enumerate}
\usepackage{amsbsy}
\usepackage{amssymb}
\usepackage{amsthm}
\usepackage{amscd}
\usepackage{float}
\usepackage{color}
\usepackage[numbers,sort&compress,square]{natbib}

\usepackage{listings}
\usepackage{stackrel}
\usepackage{authblk}
\usepackage{dsfont}
\usepackage{ulem}
\makeatletter
\newcommand{\removelatexerror}{\let\@latex@error\@gobble}
\makeatother

\usepackage{array}
\newcolumntype{L}[1]{>{\raggedright\let\newline\\\arraybackslash\hspace{0pt}}m{#1}}
\newcolumntype{C}[1]{>{\centering\let\newline\\\arraybackslash\hspace{0pt}}m{#1}}
\newcolumntype{R}[1]{>{\raggedleft\let\newline\\\arraybackslash\hspace{0pt}}m{#1}}

\begin{document}

\ifCLASSINFOpdf

\fi

\def\BibTeX{{\rm B\kern-.05em{\sc i\kern-.025em b}\kern-.08em
    T\kern-.1667em\lower.7ex\hbox{E}\kern-.125emX}}
 \setlength{\columnsep}{0.21in}

\title{Multi-Tier Hierarchical Federated Learning-assisted NTN for Intelligent IoT Services}
\author{
Amin Farajzadeh, Animesh Yadav, and Halim Yanikomeroglu
\thanks{A. Farajzadeh and H. Yanikomeroglu are with the Non-Terrestrial Networks (NTN) Lab, Department of Systems and Computer Engineering, Carleton University, Ottawa, ON K1S 5B6, Canada. A. Yadav is with the School of EECS, Ohio University, Athens, OH, 45701 USA.}}

\maketitle

\begin{abstract}
In the ever-expanding landscape of the Internet of Things (IoT), managing the intricate network of interconnected devices presents a fundamental challenge. This leads us to ask: ``\textit{What if we invite the IoT devices to collaboratively participate in real-time network management and IoT data-handling decisions?}'' This inquiry forms the foundation of our innovative approach, addressing the burgeoning complexities in IoT through the integration of non-terrestrial networks (NTN) architecture, in particular, integrated vertical
heterogeneous network (VHetNet), and multi-tier hierarchical federated learning (MT-HFL) framework. VHetNets transcend traditional network paradigms by harmonizing terrestrial and non-terrestrial elements, thus ensuring expansive connectivity and resilience, especially crucial in areas with limited terrestrial infrastructure. The incorporation of MT-HFL further revolutionizes this architecture, distributing intelligent data processing across a multi-tiered network spectrum, from edge devices on the ground to aerial platforms and satellites above. In this paper, we explore MT-HFL's role in fostering a decentralized, collaborative learning environment, enabling IoT devices to not only contribute but also make informed decisions in network management. This methodology adeptly handles the challenges posed by the non-independent and identically distributed (non-IID) nature of IoT data and efficiently curtails communication overheads prevalent in extensive IoT networks. Significantly, MT-HFL enhances data privacy, a paramount aspect in IoT ecosystems, by facilitating local data processing and limiting the sharing of model updates instead of raw data. By evaluating a case-study, our findings demonstrate that the synergistic integration of MT-HFL within VHetNets creates an intelligent network architecture that is robust, scalable, and dynamically adaptive to the ever-changing demands of IoT environments. This setup ensures efficient data handling, advanced privacy and security measures, and responsive adaptability to fluctuating network conditions. 
\end{abstract}

%
\IEEEpeerreviewmaketitle

\section{Introduction}
The Internet of Things (IoT) represents a monumental leap in the ability of humans and machines to communicate, interact, and collaborate autonomously with a myriad of devices, ranging from smart home appliances to smart cities, and from environmental surveillance systems to cutting-edge industrial machinery~\cite{iot1}. The exponential growth in IoT has brought forth challenges pertaining to extended network coverage, massive and robust connectivity, data heterogeneity, restricted storage and computing capabilities, voluminous data handling, and heightened privacy and security~\cite{iot2}. To effectively address these challenges, there is a pressing need for a robust and flexible network architecture. Herein lies the significance of the non-terrestrial networks (NTN) architecture which stands as a pivotal solution in addressing the demands of modern IoT services and networks~\cite{NTN}. NTNs, in particular, vertical heterogeneous networks (VHetNets), are capable of addressing challenges such as extended network coverage beyond terrestrial limits, offering enhanced connectivity and resilience, vital in areas where terrestrial infrastructure is insufficient or compromised~\cite{vhetnet}. 

VHetNets integrates various network layers, including terrestrial base stations, low-altitude platform stations (LAPSs), such as uncrewed aerial vehicles (UAVs), high-altitude platform stations (HAPSs), and satellites, to create a multi-tiered communication ecosystem~\cite{vhetnet2}. This integration offers several advantages for IoT networks, such as extended coverage, improved capacity, enhanced reliability, ubiquitous connectivity, and support for heterogeneous applications. However, alongside these benefits come challenges such as highly heterogeneous and expansive network architecture~\cite{Amin2}. These challenges include managing the complex interplay between different network layers, ensuring seamless connectivity across varied network platforms, and security and privacy concerns~\cite{Tasneem}.
  
In navigating these complexities and harnessing the full capabilities of VHetNets while addressing other challenges pertaining to voluminous data and their heterogeneity and privacy in IoT networks, facilitating intelligent interactions among a diverse array of IoT devices is a promising approach. Intelligent interactions not only encourage collaborative learning and decision-making but also equip the network with the ability to self-organize and evolve in response to dynamic conditions~\cite{iot3}. To embed intelligence into VHetNet architecture for IoT, we propose a federated learning (FL) framework-based multi-tier hierarchical-FL (MT-HFL) approach. Uniquely suited to IoT networks, MT-HFL excels in distributed learning across various network tiers, effectively addressing issues like data heterogeneity (i.e., non-independent and identically distributed (non-IID)) in IoT~\cite{iid}. Additionally, it manages the communication overhead prevalent in extensively connected networks~\cite{overhead}. A crucial benefit of MT-HFL, inherent to the FL paradigm, is its capacity to significantly enhance data privacy. FL allows processing data locally on IoT devices and only sharing model updates rather than raw data; MT-HFL ensures privacy across the IoT network at a lower cost of communication. This approach to learning and adaptation within VHetNets' multiple tiers is instrumental in resolving key challenges, thereby enhancing the overall functionality and efficiency of IoT networks.

Moreover, the hierarchical structure of MT-HFL enables IoT devices within the same tier to collaborate in the learning process, leading to faster convergence and more efficient data processing. Its tier-based structure also enables the exchange of knowledge among different IoT networks located in distant regions, leading to accelerated real-time responses to IoT services/demands, improved network resource management, and enhanced predictive analytics for proactive decision-making.


In this paper, we delve deep into the synergy between the integrated VHetNet architecture and MT-HFL, demonstrating why this combination is not just suitable but essential for the future of intelligent IoT infrastructures. Our contributions include a detailed exploration of the challenges inherent in VHetNets, the proposition of a novel MT-HFL framework tailored to the challenges in IoT environments, and an evaluation of its effectiveness through a simulated case study. We also outline potential use cases for this framework and suggest future research directions.

\begin{figure*} [!t]
    \centering
    \includegraphics[width=180mm]{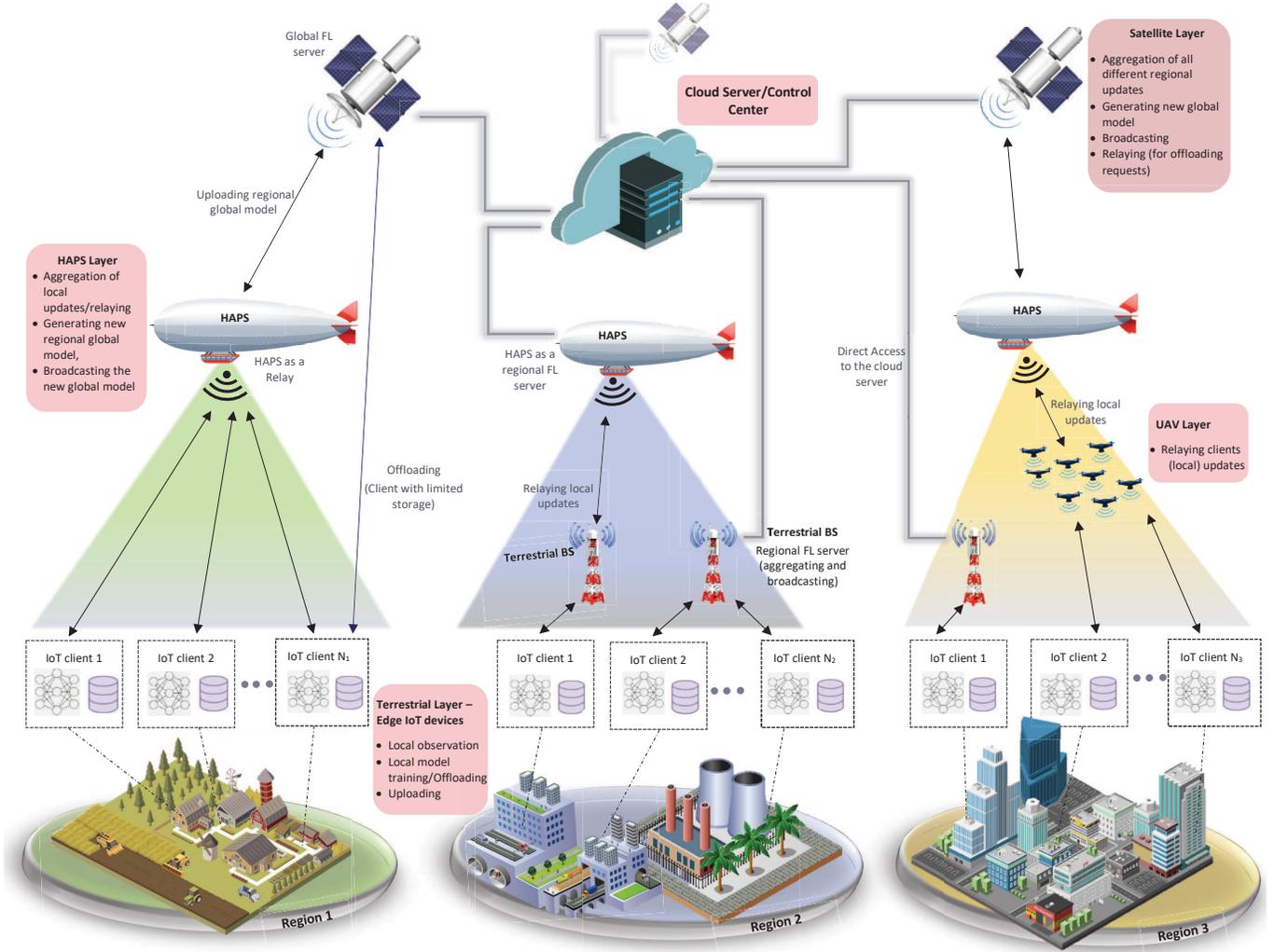}
    \caption{MT-HFL framework within VHetNet architecture in IoT environments.}
    \label{fig:sys_model}
\end{figure*}


\section{Why VHetNet Architecture for IoT Networks?}
The VHetNet architecture represents a paradigm shift in network architecture, particularly suitable for the ever-evolving number of devices, coverage, and capacity requirements of IoT systems. This section describes the inherent characteristics of this architecture and how it meets the challenging demands of IoT environments.

\subsection{VHetNet Architecture and its Characteristics}
VHetNets are formed by integrating layers of terrestrial, LAPS, HAPS, and satellite networks and hence, follow the multi-tier structure. This integration is essential in providing comprehensive coverage from the ground to the sky. VHetNets extends network coverage significantly, reaching remote, rural, and otherwise hard-to-serve areas due to the presence of aerial and satellite nodes. With its layered structure, VHetNets offers enhanced data capacity, accommodating the high volumes of data traffic typical in densely populated environments. The architecture supports a range of connectivity options, facilitating seamless communication across diverse network nodes, from traditional cellular to aerial and satellite nodes. Nodes have heterogeneous operational characteristics and capabilities. This heterogeneity allows the VHetNets to adapt its configuration to changing network conditions and user demands. By allowing load-sharing among different nodes in the network, VHentNets also offers scalability. Further, the heterogeneous and multi-layered nature of VHetNets provides a degree of inherent resilience and fault tolerance, ensuring network robustness and reliability~\cite{Amin}.

\subsection{VHetNets for IoT Networks}
The multi-tier VHetNets offer expansive and continuous coverage, a necessity for IoT devices spread across diverse geographical areas. This is crucial for ensuring consistent connectivity in both urban and remote areas, facilitating uninterrupted IoT services. As IoT networks expand, the need for scalable architecture becomes paramount. VHetNets' multi-layered approach allows for flexibility in network expansion, accommodating an increasing number of devices without compromising on performance, making it ideal for the scalable nature of IoT deployments. 

    IoT encompasses a wide range of applications, each with different connectivity and data handling requirements~\cite{iot4}. The flexible and diverse connectivity options offered by VHetNets make them suitable for a broad spectrum of IoT applications, from high-bandwidth industrial use cases to low-power environmental monitoring. Additionally, it provides robust and reliable connectivity.
     Further, to overcome the challenge of limited storage in IoT devices, the architecture integrates distributed storage options at different layers of VHetNets. The storage-intensive tasks from the devices can be offloaded to the cloud or edge nodes, enhancing overall network efficiency and data accessibility.

It is now evident from the previous discussion that VHetNets is an ideal candidate for supporting diverse IoT applications. However, both VHetNets and IoTs have challenges related to network management and massive heterogeneous data management while preserving privacy, respectively. To address these challenges jointly, we embed intelligence into VHetNets for IoT via MT-HFL framework. 
\section{Multi-Tier Hierarchical Federated Learning (MT-HFL) Framework}

\subsection{Definition of MT-HFL}
MT-HFL represents a sophisticated machine learning paradigm based on FL, specifically designed to address the challenges of VHetNets for IoT. Its defining characteristics include the following three learning aspects: 
\subsubsection{Hierarchical Learning Across Network Tiers} MT-HFL orchestrates a structured learning process across diverse network layers, each with its unique operational environment. At the edge tier, learning is tailored for real-time, data-intensive operations, while higher tiers like LAPS, HAPS, and satellites handle broader, more strategic data analysis and model refinement. The tiered approach ensures that each layer contributes optimally leveraging its computational strengths and data processing capabilities.
\subsubsection{Distributed Learning with Centralized Oversight} While the learning process is distributed, allowing local data processing and model training, there is a centralized system ensuring overall model consistency and convergence. This hybrid approach not only maximizes the benefits of localized learning but also maintains the integrity and direction of the broader network learning goals.

\subsubsection{Customized Model Learning for Each Tier} Recognizing the distinct data environments and computational capacities at each network tier, MT-HFL supports the development of tailored learning models. This customization significantly boosts the processing efficiency and accuracy of the models, catering to the specific requirements of each tier.
\subsection{Components and Functionalities of the MT-HFL Framework}

The MT-HFL framework encompasses various tiers, each with specific roles and functionalities:
\begin{itemize}
\item \textbf{Central Cloud/Control Center:} This unit acts as the central hub for coordinating the overall federated learning process, integrating insights from all tiers and disseminating refined models and decisions back through the network.
\item \textbf{Satellite Layer:} This layer acts as global FL server for global-scale coordination, which significantly expands the reach of the network, facilitating extensive model training and broadcasting. Depending on the application scenario, it can also act as a critical relay point for aggregating the regional global models or or relaying them between HAPS and the Central Cloud, ensuring seamless integration of insights from diverse regions. Moreover, it also helps the clients with limited storage to offload their computational task, i.e., local training, to the Central Cloud.

\item \textbf{HAPS Layer:} Functioning as regional FL servers, HAPS are pivotal for localized model training and broadcasting, particularly in areas where decision-making is regionally focused. They also play a key role in aggregating intermediate data and managing the flow of information to and from UAVs.

\item \textbf{UAVs Layer (or Low-Altitude Platform Station (LAPS) Layer):} These platforms act as agile relay nodes, bridging the gap between edge devices and higher network tiers. They are crucial for transmitting updates and supporting devices that are resource-constrained.

\item \textbf{Terrestrial Layer - Edge IoT Devices:} At the ground level, these devices are the frontline of data collection and initial model training, focusing on efficiency and privacy. They play a significant role in offloading heavy computational tasks to higher tiers, thereby optimizing local resource utilization.

\item \textbf{Terrestrial Layer - Terrestrial Base Stations (BSs):} These stations act as crucial relay points and local FL aggregators, managing data flow and model updates between edge devices and higher network tiers. They are key to ensuring seamless connectivity and processing within the terrestrial scope of the network.

\end{itemize}

As illustrated in Fig.~\ref{fig:sys_model}, this structured implementation of MT-HFL within IoT networks demonstrates its profound capability to enhance the efficiency and effectiveness of federated learning. Each distinct tier contributes to a cohesive and scalable learning process, tailored to the diverse and ever-evolving needs of IoT environments.
\subsection{Dynamics of the MT-HFL Framework}
The MT-HFL framework introduces a dynamic approach to learning and network management in IoT environments, characterized by the following key features and functionalities:
\begin{itemize}
    \item \textbf{Smart Client Selection:} Advanced algorithms, such as machine learning-based predictive analysis, are employed to evaluate and select client devices. These algorithms consider not just the data availability but also factors like device reliability, processing capacity, and network latency. For example, in a smart city scenario, the system might prioritize data from traffic sensors with higher accuracy and faster connectivity for real-time traffic management.
    \item \textbf{Efficient Resource Utilization:} By employing strategies like data compression and selective synchronization, the framework maximizes bandwidth and minimizes energy consumption. In a connected healthcare system, this could mean prioritizing vital patient data transmission over routine monitoring information during peak network usage.
    \item \textbf{Adaptive Learning Process:}
    The framework is equipped with dynamic algorithms capable of adjusting learning models in response to environmental changes and network feedback. For example, if a sudden change in traffic patterns is detected by edge devices, the learning model can be quickly adjusted to prioritize new data types, ensuring that the learning process stays relevant and effective.
    \item \textbf{Collaborative and Distributed Learning:}
    MT-HFL utilizes a distributed data processing model where each network tier (edge, LAPS, HAPS, satellite) processes information relevant to its operational scope and capability. The edge devices handle immediate, localized data; aerial platforms like LAPS and HAPS aggregate and preprocess data for regional insights; and satellites contribute to a macroscopic understanding. This tiered approach allows for both in-depth local learning and broader pattern recognition, enabling comprehensive network intelligence.
    \item \textbf{Scalability and System Integration:}
    MT-HFL's modular design allows for scalability, accommodating an increasing number of IoT devices. It integrates with existing network infrastructures using APIs and standardized communication protocols, ensuring compatibility with both legacy and new technologies. This scalability is key for handling the growing data volumes in IoT networks without compromising performance.
    \item \textbf{Real-time Analytics and Decision Making:}
    The framework leverages edge computing for real-time data processing. By enabling immediate data analysis and decision-making at the source (edge devices), MT-HFL facilitates rapid responses to dynamic conditions, essential in applications such as autonomous vehicle navigation where delays can be critical.
    \item \textbf{Enhanced Data Privacy and Security:}
    To protect data privacy, MT-HFL implements distributed learning techniques like federated learning, where data is processed locally and only the learning models or insights (and not the raw data) are shared across the network. This is supplemented with encryption protocols and secure multi-party computation methods to safeguard data integrity and confidentiality during transmission and processing.
\end{itemize}

In Table~\ref{tab:enh_comparison}, we present a detailed comparison between the proposed MT-HFL framework within the VHetNet architecture and the conventional HFL approaches (\cite{HFL1}, \cite{HFL2}) in terrestrial architectures, specifically tailored for IoT environments. 

\begin{table*}[htbp]
\caption{Comparison of Conventional HFL and MT-HFL in IoT Environments}
\label{tab:enh_comparison}
\centering
\begin{tabular}{L{2cm}|L{7cm}|L{7.5cm}}
\toprule %
\textbf{Feature} & \textbf{Conventional HFL (within terrestrial IoT architecture)} & \textbf{MT-HFL (within NTN IoT architecture)}\\
\hline
Hierarchical Structure & Limited - Primarily involves a single layer of clients within a confined terrestrial area. & Enhanced - Incorporates a multi-tiered structure spanning edge devices, UAVs, HAPS, satellites, and central cloud, each tier with distinct computational and communication roles.\\
\hline
Decentralized FL Architecture & Limited - Typically employs a centralized server approach with direct client-server interactions. & Advanced - Employs a decentralized, multi-server architecture across tiers, enhancing model exchange and collaboration among servers.\\
\hline
Resource Management & Uniform - Treats all devices homogeneously, potentially leading to inefficiencies and imbalances in resource usage. & Dynamic - Adopts a tier-specific approach, enabling efficient resource allocation based on the capabilities and roles of devices in each tier.\\
\hline
Scalability & Constrained - Challenges in scaling up due to reliance on a centralized architecture and limited network layers. & Highly Scalable - Capable of accommodating an exponentially growing number of IoT devices and data sets, thanks to the multi-tiered architecture.\\
\hline
Privacy and Security & Basic - Relies on traditional security measures, with potential risks in data transmission to the central server. & Enhanced - Employs advanced security protocols, including differential privacy and secure multi-party computation, to safeguard data across all tiers.\\
\hline
Global vs. Regional Learning & Limited - Focused on narrower, often local scales. & Flexible - Balances global and regional learning needs, leveraging satellites and HAPS.\\
\hline
Adaptability & Moderate - Adapts to network changes within a limited scope of the terrestrial network. & Superior - Exhibits high adaptability to diverse and evolving network conditions across various tiers, enhancing overall network robustness and reliability.\\
\hline
Accuracy & Variable - Accuracy is contingent on data distribution and device density; susceptible to noise and update inconsistencies. & Higher - Achieves improved accuracy through sophisticated resource allocation, tier-specific data processing, and leveraging diverse data sources.\\
\hline
Convergence Rate & Slower - Affected by factors like NLOS paths and channel inconsistencies, leading to prolonged training periods. & Faster - Benefits from the structured tier-based aggregation and LOS communication advantages in aerial and satellite tiers, enabling quicker model convergence.\\
\hline
Energy Efficiency & Limited - Energy constraints not centrally addressed. & Optimized - Energy-efficient operations, especially at the edge device level.\\
\hline
Network Resilience & Moderate - Vulnerable to terrestrial network issues. & Enhanced - Diverse tiers add robustness against individual network failures.\\
\hline
Data Handling Capacity & Limited - Restricted by the capacity of terrestrial networks. & Expanded - Increased data handling through distributed computing across tiers.\\
\hline
Response to Mobility & Restricted - Challenges in managing mobile clients. & Dynamic - Effectively handles mobility, especially with UAVs for transient IoT clusters.\\
\bottomrule
\end{tabular}
\end{table*}
\section{Integration of MT-HFL in VHetNets for IoT Environments}
In this section, we discuss the integration of MT-HFL within the VHetNet architecture, specifically tailored for IoT environments. This integration represents the core innovation of our approach, showcasing a novel, original, and strategically designed architecture to address the multifaceted demands/services of IoT infrastructures.

\subsection{Architectural Design and Network Composition}

\begin{itemize}
    \item \textbf{Strategic Network Layering:} The architecture strategically layers the network to optimize the MT-HFL integration. This involves a meticulous arrangement of terrestrial, aerial, and satellite networks to facilitate efficient data flow and learning processes across different tiers, crucial for comprehensive IoT coverage.

    \item \textbf{Innovative Integration of MT-HFL Components:} Core to this phase is the novel integration of MT-HFL elements within each network layer. This includes specialized federated learning nodes, custom-designed for each layer’s unique characteristics, ensuring optimal learning and data processing for IoT applications.

    \item \textbf{Seamless Inter-tier Communication Framework:} A state-of-the-art communication protocol is established to enable seamless data exchange and learning model synchronization between different network tiers, addressing the challenge of maintaining coherence in a multi-tiered IoT environment.
\end{itemize}

\subsection{MT-HFL Model Development and Optimization}

\begin{itemize}
    \item \textbf{Customized Model Development for IoT Scenarios:} The MT-HFL model is uniquely crafted with an emphasis on IoT-specific data patterns and network dynamics. This includes advanced algorithms for handling large-scale, distributed IoT data, ensuring efficient and accurate learning.

    \item \textbf{Dynamic Client Selection and Data Harmonization:} Incorporating cutting-edge techniques for client selection, the model dynamically identifies optimal nodes across tiers for participation. This is coupled with sophisticated data harmonization strategies to ensure consistency and relevance of the data being processed.
    \item \textbf{AI-Driven Model Evolution:} Leveraging artificial intelligence, the MT-HFL model continuously evolves, adapting to new IoT devices, varying data streams, and changing network conditions, showcasing the architecture’s self-evolving capability.
\end{itemize}
\subsection{Training, Adaptation, and Evolution in IoT Context}

\begin{itemize}
    \item \textbf{Progressive Training Mechanisms:} The training process employs progressive learning mechanisms, allowing the model to adaptively refine its learning based on real-time feedback from the IoT environment, ensuring relevance and accuracy in its decision-making process.

    \item \textbf{Autonomous Network Adaptation:} Integrating adaptive algorithms, the architecture autonomously adjusts to changes in IoT network topology and usage patterns, maintaining optimal performance and resource utilization.
    \item \textbf{Evolutionary Learning for Pervasive IoT Integration:} The architecture features an evolutionary learning approach, enabling it to seamlessly integrate new IoT technologies and standards, ensuring its relevance and efficacy in an ever-evolving IoT landscape.
\end{itemize}
\subsection{Model Aggregation and Federated Intelligence}
\begin{itemize}
    \item \textbf{Tier-Specific Model Aggregation:} Aggregation strategies are tailored to each network tier, ensuring that the learning models reflect the unique characteristics and requirements of different IoT segments, from high-density urban areas to remote locations.

    \item \textbf{Federated Intelligence Across Tiers:} The aggregated models contribute to a federated intelligence that spans across the network tiers. This intelligence is the culmination of distributed learning, providing a comprehensive understanding and response mechanism to diverse IoT challenges.
\end{itemize}

\subsection{Resource Allocation and Performance Optimization}

\begin{itemize}
    \item \textbf{Dynamic Resource Allocation for Enhanced Efficiency:} Utilizing advanced algorithms, the architecture dynamically allocates resources such as bandwidth and computing power, optimizing for the varying demands of IoT devices and applications.

    \item \textbf{Performance Optimization through Predictive Analytics:} Employing predictive analytics, the system proactively optimizes network performance, preempting potential bottlenecks and ensuring efficient operation across all IoT network tiers.
\end{itemize}

\subsection{Security, Privacy, and Ethical Considerations}

\begin{itemize}
    \item \textbf{Robust Security Framework for IoT Data:} The architecture incorporates a robust security framework, employing advanced encryption and secure learning protocols to protect sensitive IoT data across all network tiers.
    \item \textbf{Privacy-Preserving Techniques in Federated Learning:} Privacy-preserving mechanisms are embedded within the MT-HFL process, ensuring that individual data points remain confidential while contributing to the collective learning model.

    \item \textbf{Ethical AI and Responsible Data Usage:} Adhering to ethical AI principles, the architecture ensures responsible data usage and decision-making, upholding transparency and accountability in its AI-driven processes.
\end{itemize}

\subsection{Impact and Application in IoT Environments}
The integration of the MT-HFL framework within the VHetNet architecture is strategically designed to address the unique challenges of IoT environments, which can be justified by the following key factors:

\begin{itemize}
    \item \textbf{Handling Network Heterogeneity:} In the diverse landscape of IoT, where network environments vary greatly, MT-HFL enables effective management of this heterogeneity. It allows each network tier to develop and refine learning models specifically adapted to their operational conditions, ensuring optimal performance across the varied IoT spectrum.

    \item \textbf{Scalability and Flexibility:} The MT-HFL framework supports the dynamic growth of IoT networks, offering the necessary scalability to seamlessly integrate a growing array of devices and diverse data types. Its flexible architecture ensures that the learning process efficiently evolves with the expanding network, maintaining robustness in the face of rapid IoT proliferation.

    \item \textbf{Real-time Data Processing:} Catering to the urgent need for real-time analytics in IoT, MT-HFL facilitates immediate data processing at the edge. This localized approach significantly reduces latency, thereby accelerating response times and enhancing the overall efficiency of IoT applications
    
    \item \textbf{Optimized Learning for Regional and Global Needs:}
    MT-HFL balances the learning objectives between localized, region-specific requirements and broader, global strategies. This optimization ensures that IoT networks are responsive to immediate local environments while also benefiting from aggregated, large-scale data analysis.
    \item \textbf{Enhanced Privacy and Security:} Given the sensitive nature of IoT data, MT-HFL’s decentralized approach helps enhance data privacy and security, as data can be processed locally without needing to be transmitted over the network.
\end{itemize}
\section{IoT Use-Cases Leveraging Intelligent NTN }
The integration of VHetNets with the MT-HFL framework in IoT environments opens a wide array of real-world applications and scenarios. These demonstrate the versatility, efficiency, and adaptability of this integrated approach in addressing various challenges and enhancing capabilities in diverse sectors. This section discusses some potential scenarios.
\subsection{Smart Cities}
\subsubsection{Dynamic Traffic Light Control}
Traffic sensors and cameras at street level collect data, which is initially processed locally for immediate adjustments. Further analysis is done by UAVs, providing a wider area overview, and then by higher-tier nodes (e.g., HAPS) for city-wide traffic optimization. This tiered data processing enables traffic lights to adapt in real-time, reducing congestion based on comprehensive traffic flow analysis.
\subsubsection{Smart Public Safety with Predictive Policing}
    Surveillance data from IoT devices is locally processed at the edge level for immediate threats. UAVs and HAPS aggregate broader data trends and send them to central systems where MT-HFL analyzes patterns to predict potential safety incidents, enabling proactive policing.

\subsection{Agriculture and Environmental Monitoring}
\subsubsection{Hyper-localized Agriculture Insights} Farm sensors transmit microclimate data to nearby processing units, providing instant readings necessary for immediate farm activities. Simultaneously, broader environmental data collected by aerial and satellite systems offers insights into larger trends, allowing farmers to make informed, long-term decisions for sustainable farming.
    \subsubsection{Environmental Impact Analysis} IoT sensors collect environmental data, which is first processed locally and then aggregated by UAVs/HAPS. MT-HFL in VHetNets analyzes this data to predict environmental impacts, aiding in informed decision-making for urban planning and environmental conservation.

\subsection{Health and Telemedicine}
\subsubsection{Disease Spread Prediction in Real-Time} MT-HFL processes health data from both local (hospitals, clinics) and broader (public health sensors) sources, enabling real-time tracking and prediction of disease spread, and guiding effective public health responses.
    \subsubsection{Intelligent Health Response Systems} Patient data from IoT devices is processed at multiple levels – locally for immediate health responses and at higher network levels for long-term health trend analysis. MT-HFL enables quick, informed decision-making for healthcare interventions.

 \subsection{Industrial IoT (IIoT)}
 \subsubsection{Self-Optimizing Industrial Processes} In the VHetNet architecture, the MT-HFL framework empowers individual machines to analyze operational data instantly at the edge layer. This local data processing is then complemented by broader analytics conducted at the LAPS/HAPS layers, offering a holistic view of industrial operations. Such an integrated approach enables real-time optimization of manufacturing processes, significantly reducing downtime and enhancing overall efficiency. It ensures that each component in the industrial setup is functioning optimally, adapting swiftly to any operational changes or anomalies.
     \subsubsection{Supply Chain Synchronization} Leveraging the MT-HFL framework, real-time data from sensors along the entire supply chain undergoes multi-tiered analysis, from local processing at individual links to broader synthesis across the network. This structure allows for dynamic operational adjustments, enhancing the supply chain's overall efficiency and agility. It ensures that each segment of the supply chain can swiftly respond to changing demands, minimizing delays, and optimizing resource allocation.

\subsection{Automotive and Smart Transportation}
\subsubsection{Smart Parking Solutions}  Using the MT-HFL framework, parking sensors provide real-time occupancy data, processed locally for immediate user information. With the help hierarchical structure, this data is also synthesized at a higher layer for city-wide parking management strategies, including dynamic pricing and availability predictions.
    \subsubsection{Enhanced V2X Communications for Autonomous Vehicles} Data from vehicles and nearby roadside units is processed locally, either at terrestrial base stations or nearby LAPS, for quick navigational decisions. Beyond this, the MT-HFL framework broadens its scope by integrating traffic and environmental data from various geographical regions. This approach not only enhances the understanding of wider traffic patterns but also enables the application of effective traffic management strategies in different areas. By consolidating this diverse regional data, the framework provides advanced predictive analytics, essential for long-term traffic planning and management.
\begin{figure} [!t]
\hspace*{-0.5cm}  
    \centering 
   \includegraphics[width=100mm]{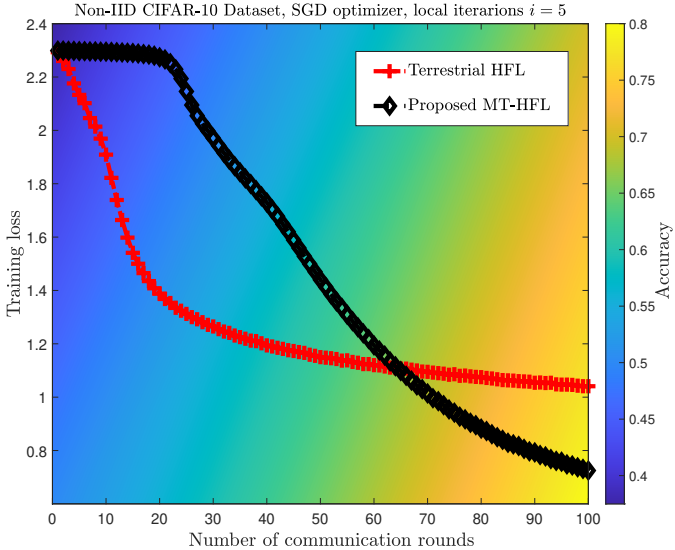}
    \caption{Training loss and accuracy evaluation versus the number of communication rounds.}
    \label{fig1}
\end{figure}
\begin{figure} [!t]
    \centering 
   \includegraphics[width=90mm]{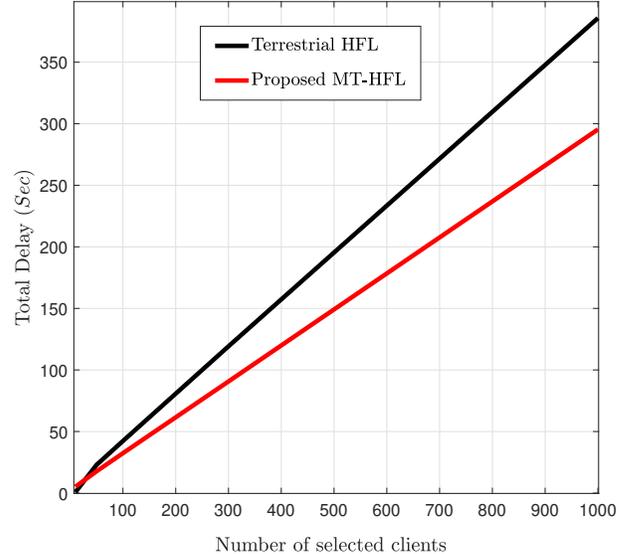}
    \caption{Total communication and computation delay performance versus the number of selected clients.}
    \label{fig2}
\end{figure}
\section{Simulation: A Case Study}
In this section, we evaluate the performance of the proposed MT-HFL framework within a simple VHetNet architecture, considering the characteristics of IoT environments. Specifically, we consider a HAPS as a server operating at an altitude of $25$ Km with a large footprint covering more than $1000$ clients on the ground. We use the non-IID CIFAR-10 dataset and stochastic gradient descent (SGD) as the optimizer in the training process with a fixed number of local iterations set at 5. To enable a fair comparison with conventional terrestrial HFL, we consider multiple terrestrial base stations with coverage of $1$ Km each, covering less than $100$ clients for the terrestrial scenario, where the channel model includes non-line-of-sight (NLOS) paths~\cite{LoS}. For VHetNets, we consider LOS paths. Furthermore, to better capture the high heterogeneity in VHetNets compared to the terrestrial scenario, we increase the level of non-IIDness of data distribution for the VHetNet dataset.

Fig.~\ref{fig1} evaluates the performance of the training loss and accuracy with respect to the number of communication rounds in the HFL process. Our results show that the proposed MT-HFL approach outperforms the conventional HFL by increasing $60$ communication rounds, as the convergence rate and accuracy substantially improve with enough clients and communication rounds. Specifically, the proposed framework achieves up to a $15\%$ improvement in performance over conventional HFL when the number of communication rounds exceeds $90$. However, in VHetNets, where the non-IIDness level of data distribution is high, inviting fewer clients and running the HFL process with fewer communication rounds lead to lower accuracy and higher training loss than in the terrestrial scenario where the non-IIDness level is not high. On the other hand, for terrestrial HFL, since the non-IIDness level is lower, the performance is better with fewer communication rounds. 

The total delay behaviour of the proposed MT-HFL and terrestrial HFL, in terms of the number of selected clients, is examined in Fig.~\ref{fig2}. The total delay includes both communication and computation delays until the convergence is achieved. Our results indicate that, by inviting the same number of clients to the learning process, the proposed MT-HFL outperforms the conventional HFL in a single-tier terrestrial scenario, achieving the same accuracy level in less time. This is primarily due to the high channel failure rate in terrestrial settings caused by NLOS paths, which often prevents clients from delivering their local updates to the server. Consequently, the coverage speed of terrestrial HFL is significantly compromised. For example, with a fixed number of $1000$ selected clients, the total delay is $25\%$ greater in terrestrial HFL compared to the proposed MT-HFL approach.

\section{Open Issues and Future Research Directions}
\begin{itemize}
    \item {\bf{Jitter-Tolerant Communication and Model Training:}} Jitter is an unwanted displacement of aerial nodes that results in communication delays, data loss, and reduced accuracy in the model training process. Specialized algorithms and techniques such as predictive modeling, advanced tracking systems, real-time network monitoring, redundancy, and error correction. These techniques can adapt to changing network conditions and mitigate the effects of jitter, ensuring the stability of communication links and uninterrupted model training in MT-HFL in VHetNets.
    \item {\bf{Learning with Unreliable Wireless Channel:}}
    The reliability of MT-HFL in VHetNets is compromised by unstable wireless channels and high client interference, causing data issues and training delays. To mitigate this, it's crucial to develop real-time adaptive strategies, including predictive modelling, error correction, redundancy, and network monitoring, along with data techniques like clustering and dimensionality reduction, ensuring model accuracy and reliability.
    \item {\bf{Information Freshness:}} In dynamic environments such as VHetNets, where network and user characteristics can change rapidly across multiple tiers, stale data can lead to inaccurate models and reduced performance. To address this challenge, techniques such as data filtering, prioritization, and dynamic data collection must be implemented to ensure that the most relevant and current data is used for model training. Additionally, advanced scheduling algorithms and real-time data processing can help to ensure that the models are continuously updated and trained with the latest available data.
\end{itemize}
\section{Conclusion}
This paper has demonstrated how the integration of the multi-tier hierarchical federated learning (MT-HFL) framework with vertical heterogeneous networks (VHetNets) innovatively addresses the specific complexities and demands of IoT environments. By tackling challenges such as extended network coverage, massive connectivity, data heterogeneity, restricted storage and computing capabilities, network reliability, and security, this integrated approach significantly enhances the efficiency and adaptability of IoT infrastructures. The MT-HFL framework, in conjunction with VHetNets, not only optimizes data handling across diverse IoT platforms but also ensures robust and extensive network coverage. Our exploration reveals that this synergistic architecture is crucial for meeting the ever-increasing data and connectivity requirements of IoT systems, thus paving the way for more intelligent, responsive, and secure IoT applications. The findings from this research provide a solid foundation for future advancements in IoT network architectures, highlighting the immense potential of this approach in a rapidly evolving technological landscape.

\vspace{-10mm}
\begin{IEEEbiographynophoto}{Amin Farajzadeh}
\textbf{[SM]} (aminfarajzadeh@sce.carleton.ca) is currently a Ph.D., student in the Department of SCE at Carleton University. 
\end{IEEEbiographynophoto}
\vskip -2\baselineskip plus -1fil
\begin{IEEEbiographynophoto}{Animesh Yadav}
\textbf{[SM]} (yadava@ohio.edu) is an Assistant Professor in the School of EECS at Ohio University, Athens, OH, USA. He is a Senior Member of IEEE.
\end{IEEEbiographynophoto}
\vskip -2\baselineskip plus -1fil
\begin{IEEEbiographynophoto}{Halim Yanikomeroglu}
\textbf{[F]} (halim@sce.carleton.ca) is a Full Professor in the
Department of SCE at Carleton University.
His research focus is on aerial and satellite networks for the
6G era. 
He is a Fellow of IEEE and a Distinguished Speaker for both IEEE
Communications Society and IEEE Vehicular Technology Society.
\end{IEEEbiographynophoto}
\end{document}